\documentclass[preprints,article,accept,oneauthor,pdftex]{Definitions/mdpi} 

\history{}

\pdfoutput=1

\usepackage{lineno}

\usepackage{amsfonts}
\usepackage{graphicx}
\usepackage{amsmath}
\usepackage{bm}
\usepackage{color}
\newcommand{\be}{\begin{equation}}
\newcommand{\ee}{\end{equation}}
\newcommand{\bwt}{\begin{widetext}}
\newcommand{\ewt}{\end{widetext}}
\newcommand{\bea}{\begin{eqnarray}}
\newcommand{\eea}{\end{eqnarray}}
\newcommand{\ket}[1]{|#1\rangle}
\newcommand{\bra}[1]{\langle #1|}

\newcommand{\ex}[1]{\langle #1\rangle}

\newcommand{\tr}{\mbox{tr}}
\newcommand{\Herm}{\mbox{Herm}}
\newcommand{\Real}{\mbox{Re}}

\newcommand\Schr{Schr\"odinger\,}

\newcommand{\Ho}{\hat H}
\newcommand{\ro}{{\hat\rho}}
\newcommand{\xo}{\hat x}
\newcommand{\po}{\hat p}
\newcommand{\xe}{\langle\xo\rangle}
\newcommand{\pe}{\langle\po\rangle}
\newcommand{\xoc}{\xo_c}
\newcommand{\poc}{\po_c}
\newcommand{\psiket}{\ket{\psi}}
\newcommand{\Aoc}{{\hat A}_c}

\newcommand{\To}{\hat T}
\newcommand{\xrm}{\mathrm{x}}

\Title{Spontaneous Wave Function Collapse with Frame Dragging and Induced Gravity}

\Author{Lajos Di\'osi}

\AuthorNames{Lajos Di\'osi}

\address[1]{%
\quad Wigner Research Centre for Physics, H-1525 Budapest 114 , P.O.Box 49, Hungary}

\corres{Correspondence: diosi.lajos@wigner.mta.hu}

\abstract{
I impose the Newtonian criteria of inertial frames on the c.o.m.
trajectories of massive objects undergoing spontaneous collapse
of their wave function. The corresponding modification of the so
far used stochastic \Schr equation eliminates the Brownian
motion of the c.o.m., restores the exact inertial motion for free
masses. For the collapse of \Schr cat states the Born rule
is satisfied invariably. The proposed machinery comes from the
radical assumption that, in the vicinity of the spontaneously
localized mass, the stochastic fluctuations of the c.o.m.  ---inevitable
in the collapse process--- would drag the physical inertial frame with
themselves. The perspective of a general theory is presented where the
spontaneous-collapse-caused breakdown of local energy-momentum 
conservation could be remedied by altering the metric, resulting in
collapse-induced curvature of the space-time.}

\begin{document}
\section{Introduction}\label{I}
Spontaneous collapse (SC) theories, reviewed in \cite{BasGhi03,Basetal13}, propose that emergence of classical
data in quantum systems is \emph{spontaneous and universal}, does not require quantum measurements. 
Spontaneous wave function collapses happen everywhere time-continuously. They recover the results of standard collapse if typical
measurement setups are considered. Models of SC fulfill two particular requirements: 
i) macroscopic degrees of freedom themselves should not develop 
superposition of macroscopically different states and ii) if such state, the so-called \Schr cat state
\bea\label{cat}
\psiket=\alpha\ket{\psi_1}+\beta\ket{\psi_2}
\eea 
with macroscopically different $\ket{\psi_1},\ket{\psi_2}$ were somehow
prepared then it should quickly get collapsed through a random process either to $\ket{\psi_1}$ or
to $\ket{\psi_2}$ with the probabilities $\vert\alpha\vert^2$, $\vert\beta\vert^2$
according to the Born-rule.

Available SC theories fulfill the above requirements at a cost of additional noise:
the dynamics of  $\psiket$ is charged by a diffusive motion in the Hilbert space
which is mandatory consequence of the SC mechanisms. This means not
only the non-conservation of momenta and energy but even an eternal increase of the latter. 

Here I am going to outline a  novel concept where SC is combined with
the redefinition of the local inertial frame in such a way that the conservation
of momentum in c.o.m. motion of massive objects is restored. 
Sec. \ref{II} recapitulates the c.o.m. dynamics
of quantized massive objects according to SC theories, Sec. \ref{III} explains
our motivations to include drag of the local inertial frame, Sec. \ref{IV} derives,
Sec. \ref{V} analyses the drag-related modifications of equations learned in Sec. \ref{II}. 
The principles of a general theory are outlined in Sec. \ref{VI}, 
predicting a possible mechanism of induced gravity. Sec. {VII} contains discussion and summary.  

\section{Spontaneous Collapse}\label{II}
SC theories are microscopic theories, slightly modifying standard non-relativistic
many-body \Schr equations in such a way that the modification remains
ignorable at atomic scales and it gets amplified for massive degrees of freedom.
For the c.o.m. wave function of a mass $M$, the \Schr equation
acquires a small but significant nonlinearity and stochasticity (exactly as if the
c.o.m. position $\xo$ were under continuous position monitoring 
\cite{Dio88a}, cf. Sec. \ref{III}, too). The corresponding stochastic \Schr equation (SSE) 
of the state $\psiket$ ---in one spatial dimension for simplicity--- reads:  
\bea\label{SSE}
d\psiket=-\frac{i}{\hbar}\Ho\psiket dt-\frac{D}{\hbar^2}\xoc^2\psiket dt
        +\frac{\sqrt{2D}}{\hbar}\xoc\psiket dW,
\eea
where $\Ho=\frac{1}{2}\po^2/M$ is the Hamiltonian and $\xoc=\xo-\xe$ 
where the brackets $\ex{.}$ will denote the expectation values in state $\psiket$. 
The stochastic term is driven by the standard Wiener process $W_t$.
The strength $D$ of the nonlinear stochastic modification depends on the
parameters of the given microscopic model of SC, is growing with $M$
but remains extreme small even for macroscopic masses. 
(Ref. \cite{Heletal17} used the precise acceleration detection of the LISA pathfinder's
$M\sim 2kg$ free test mass to put an upper bound $\sim 10^{-22}cm^2/s^3$ on $D/M^2$.)
If we consider the noise-averaged evolution, it turns out that the nonlinearities
of the SSE cancel and we are left with the linear master equation for the density matrix $\ro$: 
\bea\label{ME}
\frac{d\ro}{dt}=-\frac{i}{\hbar}[\Ho,\ro]-\frac{D}{\hbar^2}[\xo,[\xo,\ro]].
\eea
An unavoidable feature of spontaneous collapse is the eternal increase of
the average kinetic energy just like in classical \emph{diffusion} with diffusion constant $D$:
\vskip-12pt
\bea\label{egain}
\frac{d}{dt}\tr(\po^2\ro)=-\frac{D}{\hbar^2}\ex{[\xo,[\xo,\po^2]]}=2D.
\eea
The diffusive motion itself is problematic conceptually because of  momentum non-conservation. 
Solving the SSE (\ref{SSE}), the picture gets even worse. There is diffusion in both momentum
and position of the c.o.m. \cite{Dio88b}:
\vskip-24pt
\bea
d\xe&=&\frac{\pe}{M}dt+\frac{\sigma^2}{\hbar}\sqrt{8D}dW,\label{dx}\\
d\pe&=&R\sqrt{8D}dW,\label{dp}
\eea
where $\sigma^2= \ex{\xoc^2}$ and $R=\hbar^{-1}\Real\ex{\xoc\poc}$.
While momentum diffusion (Brownian motion) is a common phenomenon  classically as well,
position diffusion is not, it corresponds to discontinuity of the spatial
trajectory. But for sure, it is legitimate quantum mechanically, see our next Section.

\section{The new idea: frame-drag}\label{III}
Since SC is utilizing its analogy with standard (and time-continuous)  
collapse (cf. \cite{Dio18}), let us recollect what standard quantum theory teaches us
about standard collapse.
Consider a broad wave function and its standard collapse 
caused by a quantum measurement of position $\xo$. The measurement localizes
(collapses) the wave function at a random location, 
the localization width depends on the precision of the measurement
device. The trajectory $\{\xe_t,\pe _t\}$, let's call it
\emph{the classical trajectory}, becomes discontinuous 
at the time of collapse and the kinetic energy $\frac{1}{2}\ex{\po^2}/M$ gets 
suddenly increased. These discontinuities and non-conservation
of energy and momentum are all legitimate consequences of standard
collapse in interaction with the measurement device. In standard
theory of time-continuous measurement these features of one-shot collapse do survive.
If the c.o.m. coordinate $\xo$ is observed in a time-continuous
way \cite{Dio88a} ---monitored, in other words--- yielding the noisy \emph{signal}
\be\label{signal}
x_t=\xe_t+\frac{\hbar}{\sqrt{8D}}\frac{dW_t}{dt},
\ee
then the monitored state satisfies exactly the
SSE (\ref{SSE}) of SC where the parameter $D$ depends on the precision of the
monitoring device. In this case diffusion  of the classical trajectory $\{\xe_t,\pe _t\}$ is a real effect 
\footnote{ 
The monitored signal $x_t$ of Eq. (\ref{signal}) would define a
different trajectory from $\xe_t$. I argued earlier that the signal is the only variable
tangible for control like feedback \cite{Dio12}.  Feedback will be the 
paramount utensil in Sec. \ref{IV} to control frame-drag whereas sharp control by $x_t$ would not work:  $x_t$ is too singular in idealized Markovian monitoring.
This enforced my choice $\{\xe_t,\pe_t\}$ to define the classical trajectory.
}. 
According to pilot arguments in Sec. \ref{VII}, the well-known anomalies of
nonlinear quantum theories this time are regularly suppressed just by the SC mechanism.   
If not in its SSE (\ref{SSE}), why is SC different from collapse process under standard monitoring?
SC is proposed to happen universally, everywhere and every time, 
and \emph{without} the presence of monitoring devices. 
This makes violation of energy-momentum conservation universal, the 
diffusive classical trajectories and the constant gain of kinetic energy are also universal. 

And this \emph{universality} is exactly the point where I start from,
toward a refined concept of SC and a modified SSE, to maintain
continuity of spatial trajectory $\xe_t$ and conservation of momentum $\pe_t$ as well.

In Newtonian physics, an inertial frame is defined by the fact that in  it
free masses will move along straight lines at constant speeds.
If they do not, then we are in the wrong frame and we have to redefine the coordinates $x,y,z$. 
Now, if we take quantization of the objects into the account, 
and we assume SC, then any free mass will slowly diffuse away from what
we think in our frame to be  inertial motion.
Let us stick to the Newtonian definition and conclude that our frame is 
not the inertial one, our coordinates are not the Cartesian ones! 
Can we redefine our coordinate $x$ (and $y,z$ in the general case) 
in such a way that any free mass trajectory under SC move at constant speed 
(and along a  straight line in full 3d)?  In fact we can do so, provided the
masses are far enough from each other. Then we redefine their local frames
assuming that the diffusive (stochastic) part of the classical trajectory 
\emph{drags the local inertial frame}, i.e., in a certain vicinity of the given mass we redefine the
coordinates $x,y,z$ (cf. Fig. 1).

\begin{figure}[H]
\centering
\begin{minipage}{.30\textwidth}
\includegraphics[width=\textwidth]{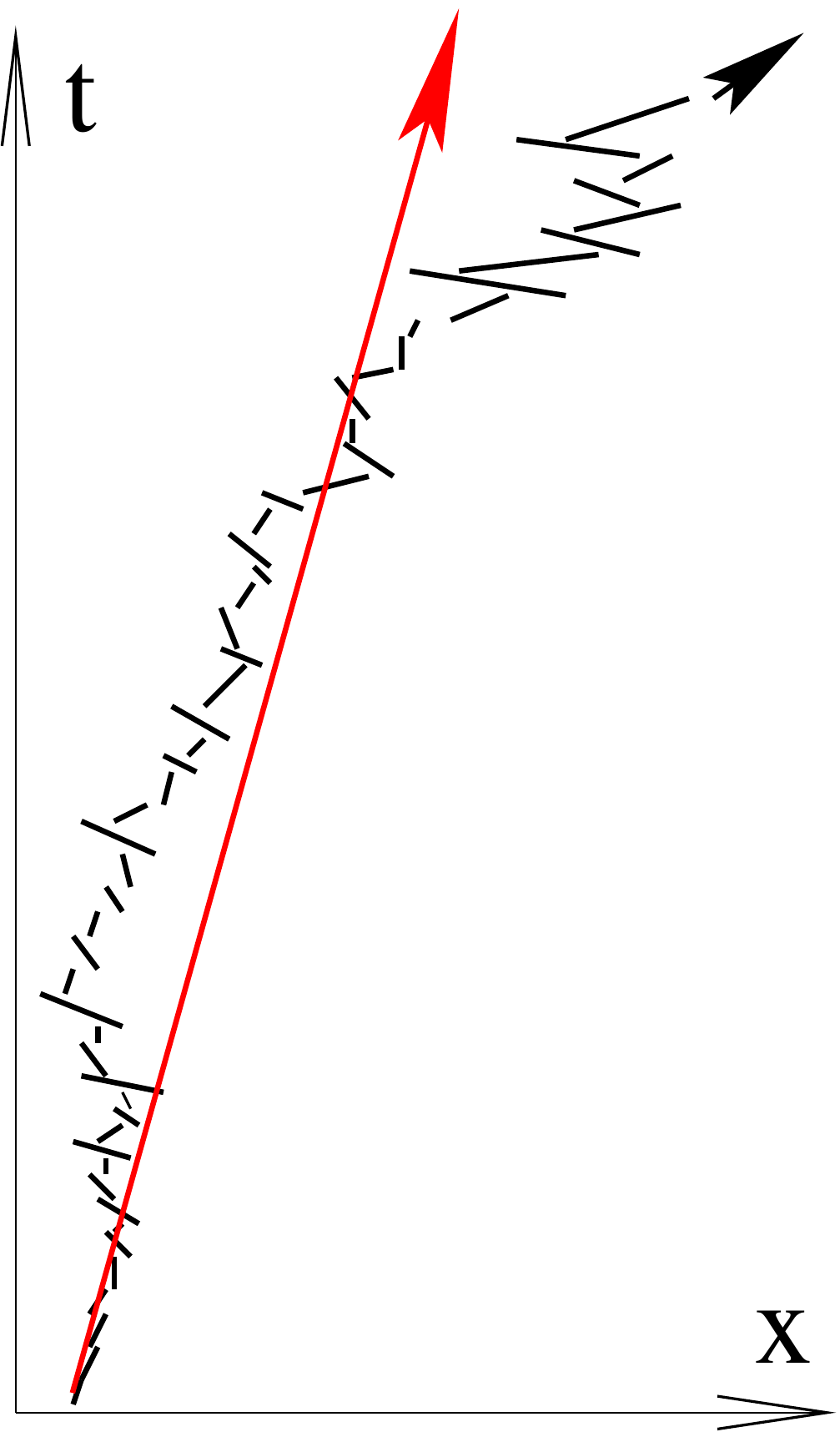}
\end{minipage}
\begin{minipage}{.32\textwidth}
\includegraphics[width=\textwidth]{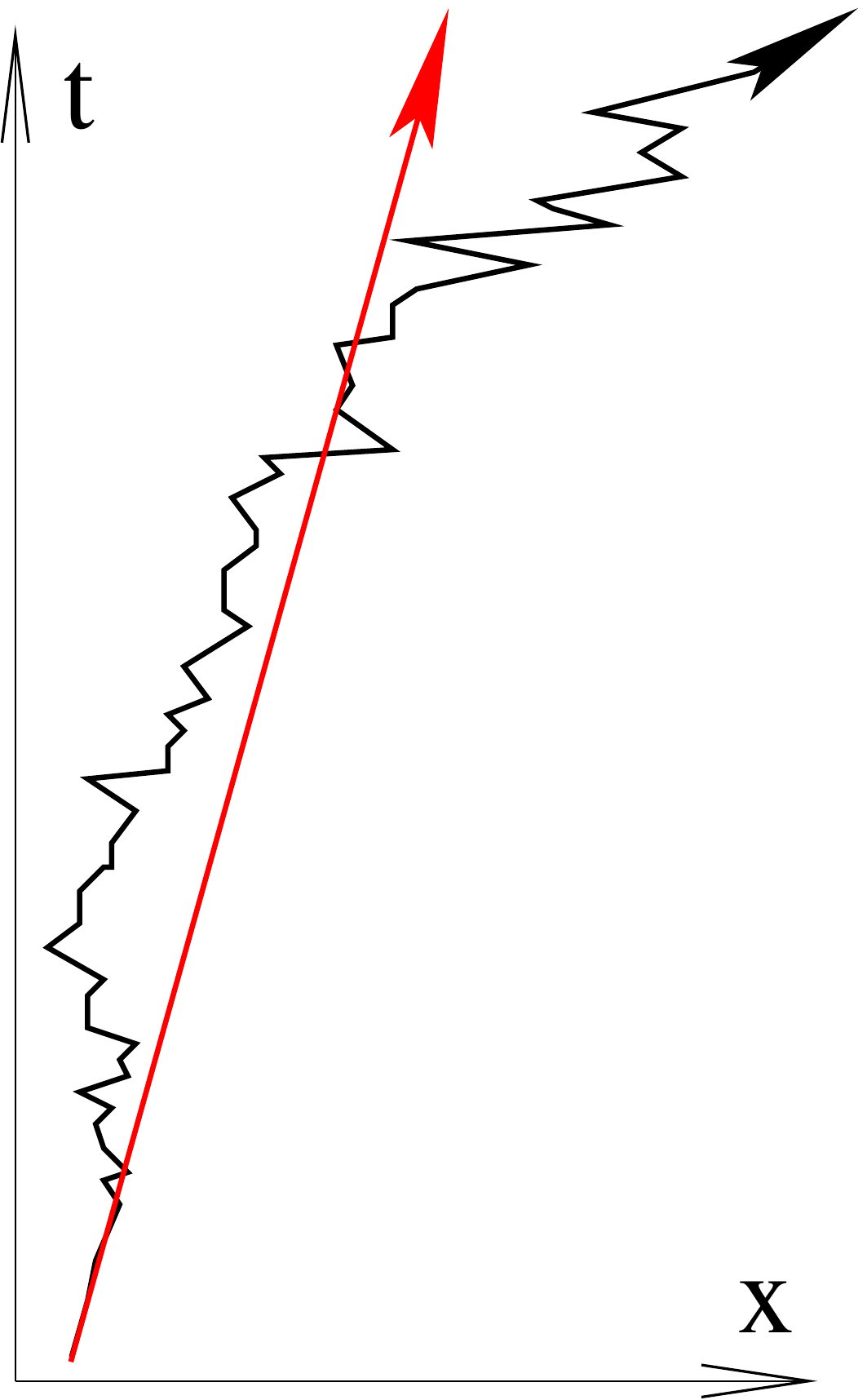}
\end{minipage}
\begin{minipage}{.30\textwidth}
\includegraphics[width=\textwidth]{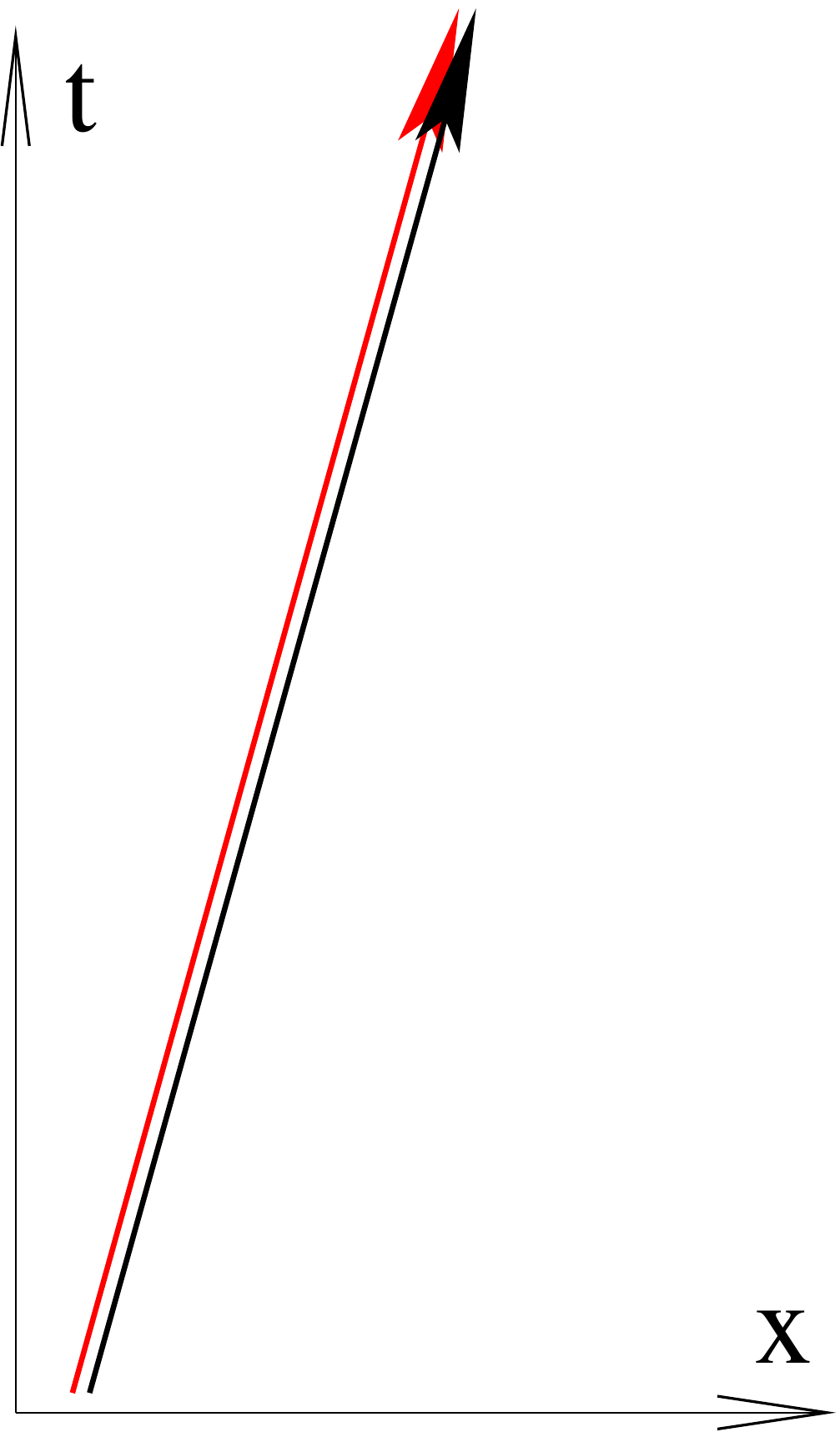}
\end{minipage}
\caption{C.o.m. trajectory of classical inertial motion (red),  
c.o.m. trajectory $\ex{\xo}_t$ of quantized motion under spontaneous collapse
(black, L), after frame-shifts (black, M), and after frame-boosts (black, R).}
\end{figure}   
 
\section{Spontaneous collapse with frame-drag}\label{IV}
We start from the SSE (\ref{SSE}) and the diffusive 
trajectory equations (\ref{dx},\ref{dp}). As said above, we assume
that the mass drags the local frame with itself to the extent
that in the new frame the diffusive part of the classical trajectory
disappears. Namely, the frame will be shifted by the distance $du$ 
and boosted by the velocity $dv$, where
\bea
du&=&vdt+\frac{\sigma^2}{\hbar}\sqrt{8D}dW,\label{dudrag}\\
dv&=&\frac{1}{M}R\sqrt{8D}dW.\label{dvdrag}
\eea 
Then in the new frame the infinitesimal evolution of the state gets two factors:
\bea
\psiket+d\psiket&=&
\exp\left\{-\frac{i}{\hbar}[R\xoc-\hbar^{-1}\sigma^2\poc]\sqrt{8D}dW\right\}\times\nonumber\\
&\times&\left\{1-\frac{i}{\hbar}\Ho dt-\frac{D}{\hbar^2}\xoc^2 dt
        +\frac{\sqrt{2D}}{\hbar}\xoc dW\right\}\psiket.
\eea
The SSE (\ref{SSE}) evolves the initial state $\psiket$ first, and the unitary factor
of frame-drag (\ref{dudrag},\ref{dvdrag}) evolves it subsequently (see same 
calculation already in Ref. \cite{Dio88b}). The resulting SSE  reads
\bea\label{SSEdrag}
d\psiket=-\frac{i}{\hbar}\Ho\psiket dt
&-&\frac{D}{\hbar^2}\left(\xoc^2 
                                                  +4[R\xoc-\hbar^{-1}\sigma^2\poc]^2
                                                  +4i[R\xoc-\hbar^{-1}\sigma^2\poc]\xoc\right)\psiket dt\nonumber\\
        &+&\frac{\sqrt{2D}}{\hbar}(\xoc-2i[R\xoc-\hbar^{-1}\sigma^2\poc])\psiket dW.
\eea
Comparing it with the old SSE (\ref{SSE}), we recognize the new terms
coming from the frame-drag. The equation can be rewritten into a compact form 
\bea
\label{SSEdragcomp}
d\psiket&=&-\frac{i}{\hbar}\left(\Ho+\Ho_\psi\right)\psiket dt
-\frac{D}{\hbar^2}\Aoc^\dagger \Aoc \psiket dt
  +\frac{\sqrt{2D}}{\hbar}\Aoc\psiket dW,\\
\label{Hdrag}
\Ho_\psi&=&\frac{4D}{\hbar}(R\xoc^2-\sigma^2{\hat R}),\\
\label{Lind}
\Aoc&=&\xoc-2i[R\xoc-\hbar^{-1}\sigma^2\poc].
\eea
With notation $\hat R=\hbar^{-1}\Herm\xoc\poc$, we defined 
the nonlinear Hamiltonian $\Ho_\psi$ and introduced the ``Lindbladian'' $\Aoc$.
The frame-drag induced a new Hamiltonian (\ref{Hdrag}) 
proportional to the small diffusion parameter $D$. When $R>0$ and
$\sigma^2$ is not too large, the induced Hamiltonian $\Ho_\psi$ is 
gauge-equivalent with a self-attracting harmonic potential
responsible for localization around $\xe$.

Whether a master equation, a modification of (\ref{ME}) exists or it does not?
If the initial state is $\ro=\psiket\bra{\psi}$ then we have
\bea\label{MEnonlin}
\frac{d\ro}{dt}
=-\frac{i}{\hbar}[\Ho+\Ho_\psi,\ro]
  +\frac{2D}{\hbar^2}\left(\Aoc\ro\Aoc^\dagger-\Herm\Aoc^\dagger\Aoc\ro\right).
\eea
This looks like a dissipative Lindblad master equation. 
But, most importantly, this is only valid at the initial time when $\ro$ is pure,
because $\Ho_\psi$ and $\Aoc$ depend on the pure state $\psiket$
and not on the average $\ro$. The loss of the closed linear evolution
of the density matrix is a warning that SC with frame-drag leads to 
an \emph{essentially nonlinear} quantum theory, and this nonlinearity opens
the door to major \emph{anomalies} like non-physical action-at-a-distance  \cite{Gis90}
and the breakdown of the Born---von-Neumann statistical interpretation of 
$\psiket$, cf. \cite{Dio16} and references therein. 
We come back to this problem later in Sec. \ref{VII}.

\section{Dynamics in the dragged frames}\label{V}
To discuss the departure of the new SSE (\ref{SSEdragcomp}) 
from the old one (\ref{SSE}) we follow two methods. First, we directly study
the new one in Appendix A, to find that
\bea
\frac{d}{dt}\tr(\po^2\ro)&=&2D(1-4R^2),\label{egaindrag}\\
d\xe&=&\frac{\pe}{M}dt,\label{dxdrag}\\
d\pe&=&0.\label{dpdrag}
\eea
The kinetic energy gain or loss in Eq. (\ref{egaindrag}) depends on the position-momentum 
correlation $R$ which is zero for a real valued wave function when
the rate of gain is $2D$ as in the SC without the drag, cf. Eq. (\ref{egain}). 
Below we see that $R$ tends asymptotically to $1/2$  and the kinetic energy will change no more. 
The Eqs. (\ref{dxdrag},\ref{dpdrag}) are counterparts of Eqs. (\ref{dx},\ref{dp}) 
--- with no stochastic terms this time. It is straightforward to see that in the
presence of a potential $V(\xo)$, our repeated derivation would modify Eq.
(\ref{dpdrag}) for $d\pe=-\ex{V^\prime(\xo)}$. This completes our statement:
the classical trajectory $\{\xe_t,\pe _t\}$ corresponds to classical motion (\`{a} la Ehrenfest).

Rather than to further discuss the SSE (\ref{SSEdragcomp}) itself, it is more
convenient to consider the SSE (\ref{SSE}) and invoke its known
features that survive the drag invariably.
According to  (\ref{SSE}), a broad initial wave packet starts to shrink
immediately, shrinking has a random component. At the same time,
the classical trajectory is charged by diffusive random fluctuations. 
Now, in the dragged frame, the process of shrinking of the wave packet 
remains the same, but the diffusive fluctuations of the trajectory go away,
see  Eqs. (\ref{dxdrag},\ref{dpdrag}).
It is known about the SSE (\ref{SSE}) \cite{Dio88b} that the solutions, after a
transient period, converge asymptotically to the Gaussian soliton 
\be
\psi_t(x)=(2\pi\sigma_\infty^2)^{-1/4} \exp\left(-(1-i)\frac{(x-\xe_t)^2}{4\sigma_\infty^2}+\frac{i}{\hbar}\pe_t x\right)
\ee
at constant spread and correlation:
\be
\sigma_\infty^2=\sqrt{\frac{\hbar^3}{8DM}},~~~~~~~~~R_\infty=1/2.
\ee
With the old  SSE (\ref{SSE}), the c.o.m. of the soliton remains subject of the phase-space diffusion (\ref{dx},\ref{dp}): 
\bea
d\xe&=&\frac{\pe}{M}dt+\sqrt{\frac{\hbar}{m}}dW,\\
d\pe&=&\sqrt{2D}dW,
\eea
but this time the coefficients are constants. In the dragged frames, i.e. with SSE (\ref{SSEdragcomp}),
this diffusion disappears. After a transient period, the free
mass wave packet becomes a soliton (cf. also Appendix A) 
and will move classically, like traveling free solitons used to.

If the initial state were a \Schr cat (\ref{cat}), where $\psiket_1$ and
$\psiket_2$ are distant (suppose standing) wave packets then the SSE (\ref{SSE}) predicts
a quick collapse of the cat into one of the components, randomly
and with the Born probabilities, respectively. Obviously the collapse
itself happens the same way in the dragged frames as well. The only
difference is that the c.o.m. will be preserved. E.g., if 
$x_1=\xe_1$ and $x_2=\xe_1$ stand for the respective quantum
expectation values of $\xo$ in the two distant wave packets in question,
then $\xe=\vert\alpha\vert^2 x_1+\vert\beta\vert^2 x_2$ is the
c.o.m. of the \Schr~cat and if the quick collapse happens, say, to $\psiket_1$ then it
gets shifted from $x_1$ to the middle, i.e. to $\xe$.
(Same happens with traveling wave packets $\psiket_1,\psiket_2$, in the velocity space.)

We conclude that the particular requirements i)-ii) for SC, mentioned in Sec. \ref{II},
that are fulfilled by the available theories, will remain satisfied by the new
SC model with frame-drag, with the bonus that, unlike in the old SC models, 
the classical trajectory $\{\xe_t,\pe _t\}$ follows the classical equations of motion, 
also conservation of momentum $\pe_t$ and the continuity of the path $\xe_t$ 
become restored. Let us remark that
we would try dragging the local clock-time too, with a 
desire if the mean kinetic energy would be conserved even in the
transient period. Next Section outlines the principles of  local energy-momentum 
conservation by local dragging $x,y,z,t$ and  with possible ``dragging'' the local metric, too.

\section{Principles of a general theory: induced gravity}
\label{VI}
As I said in Sec. \ref{II}, available SC models are microscopic and in terms of fields. 
Local frame-drags anticipate the necessity to use curvilinear coordinates.
Dynamics in curvilinear coordinates are, for historic reasons, less common non-relativistically than
relativistically. Hence I found it more convenient to consider a possible
relativistic formalism of frame-drags first. 
Harnessed with the standard theory of metric in general coordinates,
the notion of local inertial frames in the vicinity of  distant masses, used in Secs.
\ref{III}-\ref{IV}, becomes trivial. Also tailoring of distant local inertial frames
is straightforward mathematically. Standard collapses are not invariant relativistically, 
thus a relativistic concept of SC is far from being established. Nonetheless,
here I assume that there can be a related theory just for the sake of using
the relativistic formalism of space-time structure. 

Consider a flat space-time with Minkowski coordinates  $\xrm^a$ and metric $\eta_{ab}$, 
which is the background for quantized relativistic fields of matter. 
The definition of a Minkowski inertial frame, whose non-relativistic limit is Newton's
definition in Sec. \ref{III}, can be formulated in terms
of the energy-momentum tensor $T_{ab}$, by requiring that its divergence vanish:
\be\label{Mink}
\nabla_b T_a^b=0.
\ee
This is to be satisfied by $\To_{ab}$ after quantization of the relativistic fields.   
To think about relativistic SC theories, according to concepts
in \cite{TilDio16} and earlier references therein, now we suppose  the presence 
of a certain universal and spontaneous monitoring of the  energy-momentum tensor 
$\To_{ab}(\xrm)$ yielding the signal
\be\label{signalT} 
T_{ab}(\xrm)=\ex{\To_{ab}(\xrm)}+\delta T_{ab}(\xrm),
\ee
where $\ex{.}$ stands for expectation value in the Heisenberg picture
and $\delta T_{ab}(\xrm)$ is a classical (colored) noise. 
Monitoring modifies the standard unitary evolution of the fields,
the expectation values  $\ex{T_{ab}(\xrm)}$ themselves
have stochastic components. The divergence $\nabla_b\To_a^b$ does no more
vanish, neither does it so in mean:
\be\label{noconserv}
\nabla_b\ex{\To_a^b}\neq0.
\ee
We proceed in the spirit of Sec. \ref{III} and decide that 
$\xrm^a$ are not the correct Minkowski coordinates. Similarly to what we
did in Sec. \ref{IV}, we could try frame-drag locally this time:  
\be\label{dragu}
\xrm^a\rightarrow\xrm^a+\xi^a(\xrm),
\ee
to \emph{ensure $\nabla_b\ex{\To_a^b(\xrm)}=0$ in the dragged coordinates}.
(In parenthesis, we mention an alternative: we could try to construct frame
drag to ensure  that the signal (\ref{signalT}) satisfy $\nabla_b T_a^b=0$.) 
There is no guarantee that frame-drag gives us sufficient freedom to get rid
of the divergence of $\ex{\To_{ab}}$. Fortunately, we can replace
frame-drag by something more general, more plausible, and more convenient. 

The concept in Sec. \ref{III} says that the original Cartesian flat (here Minkowski) background metric 
is valid in the dragged coordinates. But we know that change of coordinates
 (\ref{dragu}) at retaining the metric $\eta_{ab}$ is equivalent to retaining the coordinates 
 and changing the metric
 \footnote{This is, in principle, valid for the global frame-drag (\ref{dudrag}) in Sec. \ref{IV} 
 yielding $g_{01}=du/dt+h.o.t.$ which is highly singular due to the idealized
Markovian model, and its interpretation is non-trivial if possible at all.}
: 
\be\label{gabdrag}
\eta_{ab}\rightarrow \eta_{ab}-\nabla_b\xi_a-\nabla_a\xi_b+
\mbox{higher order terms in } \xi.
\ee
And here we come to a turning point. 
We forget frame-drag (\ref{gabdrag}) and allow for a general change of the metric:
\be
\eta_{ab}\rightarrow \eta_{ab}+\delta g_{ab},
\ee
to \emph{ensure  
\be\label{conserv}
\nabla_b^{covariant}\ex{\To_a^b(\xrm)}=0
\ee
 in the original coordinates}. This gives us a much wider freedom, compared
to frame-drag, to solve the task. We can change the space-time's physical structure that
frame-drags leave unchanged. If violation of energy-momentum conservation by SC
goes away in a modified space-time structure then I talk about {\em gravity induced
by collapses}. 

There is no apriori guarantee that change of the metric can restore energy-momentum 
conservation (\ref{conserv}) lost by SC. Even if this were the case, the induced 
space-time curvature may be different from Einstein's that is sourced by masses,
not by wave function collapses.  The concept points anyway towards a new  coupling 
between quantized matter and classical space-time --- different from all previous high level
proposals like the semiclassical \cite{Mol62,Ros63}, the Bohmian \cite{Str15,Lal19}, 
or the stochastic semiclassical \cite{TilDio16,TilDio17} couplings.

 \section{Difficulties,  perspectives, summary}\label{VII}
The essential nonlinearity, like that of our new SC proposal (cf. Sec. \ref{III}), 
is well-feared because of major anomalies it induces.
These anomalies, on the other hand, require particular entangled states 
between distant systems. In our case, entanglement of the c.o.m. degrees of
freedom of distant \emph{massive} objects is required, otherwise the anomalies do not surface\footnote{
Instances of anomalies, fake action-at-a-distance among others,
assume an entangled composite system AB of remote parts A and B, 
and standard collapse in B where B can be a single ``massless'' spin-half system. 
In SC theories, collapse of the spin does not happen unless it gets entangled with a massive system C, 
which means A has to be entangled with a \emph{massive} system BC. 
System $A$, too, must be a massive system otherwise SC has ignorable effect on it.
}. 
This offers a loophole, specific to SC theory: such states are heavily suppressed by the very mechanism of SC.

Leading SC models are microscopic, like the DP model \cite{Dio89,Pen96} is
(though cautious Roger Penrose advocates his  ``minimalist approach''
without constructing detailed dynamics)  and like the CSL model \cite{GhiPeaRim90} is. 
In the present proposal we discussed and modified the effective SC of the 
simplest massive macroscopic d.o.f. which are the c.o.m. of massive
objects. It is not trivial how we can implement the concept of frame-drag and 
induced gravity at the microscopic scales. A creative approach can start from the
elementary mechanism  shown in Secs. \ref{III}-\ref{IV}. Alternative starting
point can be the Newtonian limit of the general theory drafted in Sec.  \ref{VI}.
A systematic test of both frame-drag and  induced gravity on DP model  \cite{inprep}
would enlighten whether the new concept is new physics or a deadlock. 

The intrinsic ---non-dynamical--- combination between SC and gravity has
been basic for the DP-model from its conception \cite{Dio89,Pen96}. 
Much later, this has been shown to be the unique combination of 
SC and semiclassical gravity that eliminates the essential nonlinearity 
of semiclassical gravity \cite{TilDio16,TilDio17}.
But the derivation of an \emph{emergent} Newtonian gravity from SC
lacked the sufficient inspiration. The precursor of the
present idea of restoring momentum-conservation and geodetic (inertial) 
motion of free masses under SC occurred ten years ago, led me to 
a sophisticated argument ---with a bit of wishful thinking--- for induced
Newtonian gravity \cite{Dio09}.  The present proposal of frame-drag,
visioned from the Newtonian definition of inertial frames, 
points towards a new relationship between SC and gravity.
While SC had been conceived as a model of emergence of classicality
in a quantized Universe, it may be responsible for the emergence of gravity as well,
as  the title of Ref. \cite{Dio09} proposed literally. 

In this work, we eliminate the well-known violation of classical equations of motion 
in the c.o.m. dynamics of masses under spontaneous wave funtion collapse,
imposing suitable transformation of the reference frame coordinates, called frame-drag.
We also outlined a general theory where the violation of the local 
energy-momentum conservation by spontaneous collapses can be 
removed by the suitable change of the space-time metric structure, 
contributing to a new concept of inducing gravity by collapses instead 
of sourcing it by masses. Relevance of this latter perspective has still to be 
confirmed or rejected by testing it on currently known spontaneous 
collapse models.  


\funding{This research was funded by 
the National Research Development and Innovation Office of Hungary Projects
Nos. 2017-1.2.1-NKP-2017-00001 and K12435, by EU COST Action CA15220, 
by the Foundational Questions Institute mini-grant.}

\acknowledgments{I appreciate many useful discussions with Thomas Konrad.}

\appendixtitles{yes} 
\appendix
\section{}
Selected features of the SSE (\ref{SSEdragcomp}) 
---with definitions (\ref{Hdrag},\ref{Lind}), and $\Ho=\frac{1}{2}\po^2/M)$--- are derived below.

 \emph{Proof of Eq. (\ref{egaindrag}).}
The SSE (\ref{SSEdragcomp}) or, equivalently, the master equation
(\ref{MEnonlin}) yields the following equation for the increment of 
$\tr(\po^2\ro)$:
\bea
d\tr(\po^2\ro)=
\left\langle
\frac{i}{\hbar}[\Ho_\psi,\po^2]dt
+\frac{D}{\hbar^2}(2\Aoc^\dagger\po^2\Aoc-\{\Aoc^\dagger\Aoc,\po^2\})dt
\right\rangle.
\eea
Work out the relationships:
\bea
{[\Ho_\psi,\po^2]}&=&8iD(R\{\xoc,\po\}-\hbar^{-1}\sigma^2\poc\po),\\
2\Aoc^\dagger\po^2\Aoc-\{\Aoc^\dagger\Aoc,\po^2\}&=&
2\hbar^2(1+4R^2)-8\sigma^2\poc\po,
\eea
and insert them:
\bea
d\tr(\po^2\ro)=
\left\langle
\frac{i}{\hbar}8iD(R\{\xoc,\po\}-\hbar^{-1}\sigma^2\poc\po)
+\frac{D}{\hbar^2}2\hbar^2(1+4R^2)-8\sigma^2\poc\po
\right\rangle=2D(1-4R^2),
\eea
where we used the identity $\ex{\{\xoc,\po\}}=2\hbar R$. 

\emph{Proof of Eq. (\ref{dxdrag}).}
The SSE (\ref{SSEdragcomp}) yields the following equation for the spatial trajectory:
\bea
d\xe=
\left\langle
\frac{i}{\hbar}[\Ho+\Ho_\psi,\xo]dt
+\frac{D}{\hbar^2}(2\Aoc^\dagger\xo\Aoc-\{\Aoc^\dagger\Aoc,\xo\})dt
+\frac{\sqrt{2D}}{\hbar}(\xo\Aoc+\Aoc^\dagger\xo)dW
\right\rangle.
\eea
The free Hamiltonian $\Ho$ yields $(\pe/M)dt$, the other terms yield zero, since
\bea
\ex{[\Ho_\psi,\xo]}&=&8iD\ex{\sigma^2\poc}=0,\\
\ex{2\Aoc^\dagger\xo\Aoc+\{\Aoc^\dagger\Aoc,\xo\}}&=&\ex{\Aoc^\dagger[\xo,\Aoc]+H.C.}
=-2\sigma^2\ex{\Aoc^\dagger}+C.C.=0,\\
\ex{\xo\Aoc+\Aoc^\dagger\xo}&=&2\ex{\xo\xoc}-2\sigma^2=0.
\eea

\emph{Proof of Eq. (\ref{dpdrag}).}
The SSE (\ref{SSEdragcomp}) yields the following equation for the momentum:
\bea
d\pe=
\left\langle
\frac{i}{\hbar}[\Ho_\psi,\po]dt
+\frac{D}{\hbar^2}(2\Aoc^\dagger\po\Aoc-\{\Aoc^\dagger\Aoc,\po\})dt
+\frac{\sqrt{2D}}{\hbar}(\po\Aoc+\Aoc^\dagger\po)dW
\right\rangle.
\eea
The r.h.s. vanishes since
\bea
\ex{[\Ho_\psi,\po]}&=&8iD\ex{R\poc-\sigma^2\xoc}=0,\\
\ex{2\Aoc^\dagger\po\Aoc+\{\Aoc^\dagger\Aoc,\po\}}&=&\ex{\Aoc^\dagger[\po,\Aoc]+H.C.}
=-i\hbar(1-2iR)\ex{\Aoc^\dagger}+C.C.=0,\\
\ex{\po\Aoc+\Aoc^\dagger\po}&=&\ex{\{\po,\xoc\}}-2\hbar R=0.
\eea

\emph{Proof of static soliton solution:} 
 \vskip-12pt
 \be
\langle x\psiket=(2\pi\sigma_\infty^2)^{-1/4} \exp\left(-(1-i)\frac{x^2}{4\sigma_\infty^2}\right).
\ee
Applying the SSE (\ref{SSEdragcomp}) to it, the stochastic term vanishes since
\be
\Aoc\psiket=i(1-2R_\infty)\psiket=0.
\ee
The yield of the deterministic terms reads:
\bea
(\Ho+\Ho_\psi)\psiket
&=&\left\{\frac{\hbar^2}{2M\sigma_\infty^2}
-i\left(\frac{\hbar^2}{4M\sigma_\infty^2}-\frac{2D\sigma_\infty^2}{\hbar}\right)
+\frac{2D}{\hbar}(2R_\infty-1)\xo^2
+i\left(\frac{\hbar^2}{4M\sigma_\infty^4}-\frac{4D}{2\hbar}\right)\xo^2\right\}
\psiket\nonumber\\
&=&\frac{\hbar^2}{2M\sigma_\infty^2}\psiket.
\eea

\reftitle{References}

\end{document}